\documentclass{article}
\usepackage[utf8]{inputenc}
\usepackage{amsmath,amssymb,amsfonts,stmaryrd}
\usepackage{physics}
\usepackage{dsfont}
\usepackage{graphicx}

\usepackage{color}
\definecolor{purple}{rgb}{0.5,0,0.5}

\usepackage{subcaption}
\usepackage{color}
\usepackage{xcolor}

\usepackage{tikz}
\usetikzlibrary{arrows,shapes,trees}
\usetikzlibrary{positioning,calc}
\usetikzlibrary{decorations.markings}

\newcommand{\ra}{\rightarrow}

\newcommand{\halpha}{{\hat\alpha}}
\newcommand{\hg}{{\hat g}}
\newcommand{\Cl}{{\rm Cl}}
\newcommand{\Irr}{{\rm Irr}}

\newcommand{\CC}{{\mathbb C}}
\newcommand{\ZZ}{{\mathbb Z}}
\newcommand{\NN}{{\mathbb N}}
\newcommand{\RR}{{\mathbb R}}

\newcommand{\HH}{{\mathbb H}}

\newcommand{\hG}{{\hat G}}

\newcommand{\cA}{\mathcal A}
\newcommand{\cB}{\mathcal B}
\newcommand{\cD}{\mathcal D}

\newcommand{\cC}{\mathcal C}

\newcommand{\cR}{\mathcal R}

\newcommand{\eps}{\varepsilon}
\newcommand{\dimr}{{{\rm dim}\, r}}

\newcommand{\vslash}{{v\!\!\!\slash}}

\usepackage{xfrac}

\usepackage[left=1in,right=1in,top=1in,bottom=1in]{geometry}

\usepackage{hyperref}

\hypersetup{colorlinks=true,urlcolor=[rgb]{0,0,0.5},citecolor=[rgb]{0,0.5,0},linkcolor=[rgb]{0,0.5,0}}

\newcommand{\hR}{{\widehat R}}
\begin{document}

\title{Free and Interacting Short-Range Entangled Phases of Fermions: Beyond the Ten-Fold Way}

\author{Yu-An Chen, Anton Kapustin, Alex Turzillo, and Minyoung You\\California Institute of Technology, Pasadena, CA 91125}

\maketitle

\begin{abstract}
We extend the periodic table of phases of free fermions in the ten-fold way symmetry classes to a classification of free fermionic phases protected by an arbitrary on-site unitary symmetry $\hat G$ in an arbitrary dimension. The classification is described as a function of the real representation theory of $\hat G$ and the data of the original periodic table. We also systematically study in low dimensions the relationship between the free invariants and the invariants of short-range entangled interacting phases of fermions. Namely we determine whether a given symmetry protected phase of free fermions is destabilized by sufficiently strong interactions or it remains stable even in the presence of interactions. We also determine which interacting fermionic phases cannot be realized by free fermions. Examples of both destabilized free phases and intrinsically interacting phases are common in all dimensions. 

\end{abstract}

\section{Introduction}

It is well-known by now that short-range-entangled (SRE) phases of free fermions on a lattice can be classified using K-theory \cite{Kitaev}, or equivalently using the topology of symmetric spaces \cite{classification,tenfold}. Originally, the classification was done in the framework of the ten-fold way, where the only allowed symmetries are charge conservation, time-reversal, particle-hole symmetry, or a combination thereof. But the K-theory framework can also be extended to systems with more general symmetries, both on-site and crystallographic \cite{TeoFuKane,MongEssinMoore,liangfu,KdBvWKS,FreedMoore,andofu}. The answer is encoded in an abelian group, with the group operation corresponding to the stacking of phases.

The first goal of this paper is to derive the classification of free fermionic SRE phases with a unitary on-site symmetry $\hat G$ in arbitrary dimensions. We show that in any dimension representation-theoretic considerations reduce the problem to classifying systems of class D, A, and C. Since the solution of the latter problem is well-known, the key step in the derivation is the reduction from a general symmetry $\hat G$ to ten-fold symmetry classes. Such a reduction is not new and has been described in detail in Ref. \cite{HHZ}. But since the authors of \cite{HHZ} work with complex fermions, and for our purposes it is more convenient to use Majorana fermions, we give a new proof of the reduction. The classification is described succinctly in Table \ref{table:G-classif}. We have included in Table \ref{table:examples} the results of applying this general classification formula to some common symmetry groups.

When we consider systems with symmetries other than the ten-fold way symmetries, it is no longer useful to adopt the ten-fold way nomenclature. For example, a fermionic system with a $U(1)\times G$ symmetry, where the generator of $U(1)$ is the fermion number, can equally well be regarded as a symmetry-enriched class A system and as a symmetry-enriched class D system. On the other hand, the distinction between unitary and anti-unitary symmetries remains important. If we denote by $\hat G$ the total symmetry group (including the fermion parity $\ZZ_2^F$), this information is encoded in a homomorphism
\begin{equation}
\rho:\hG\ra\ZZ_2^T.
\end{equation}
We also need to specify an element $P\in\hG$ which generates the subgroup $\ZZ_2^F$. This elements satisfies $P^2=1$ and is central.\footnote{Centrality is equivalent to the assumption that all symmetries are bosonic, i.e. do not exchange bosons with fermions.} Since $P$ is unitary, we must have $\rho(P)=1$ (here we identify $\ZZ_2$ with the set $\{1,-1\}$). 
The symmetry of a fermionic system is encoded in a triplet $(\hG,P,\rho)$. For example, class D systems correspond to a triplet $(\ZZ_2,-1,\rho_0)$, where $\rho_0$ is the trivial homomorphism (sends the whole $\hG$ to the identity), while class A systems correspond to a triplet $(U(1),-1,\rho_0)$. In this paper we study only systems with unitary symmetries, i.e. we always set $\rho=\rho_0$. We allow $\hG$ to be an arbitrary compact Lie group, with the exception of section 3.3, where $\hG$ is assumed to be finite.

The reader might notice that many of our results on the classification of free systems can be naturally expressed in terms of equivariant K-theory. The connection between free systems with an arbitrary (not necessarily on-site or unitary) symmetry and equivariant K-theory has been studied in detail in \cite{FreedMoore}. However, in this paper we prefer to use more elementary methods, such as representation theory of compact groups. This has the advantage of making clear the physical meaning of K-theory invariants, which is crucial for the purpose of comparison with interacting systems.

The second goal of this paper is to study the relationship between the free classification and the classification of short-range entangled interacting fermionic phases. When interactions of arbitrary strength are allowed, the classification of SRE phases of fermions is much more complicated, but in low dimensions\footnote{An answer in an arbitrary number of dimensions was conjectured in \cite{Kapustin2015}.} the answer is known for an arbitrary finite on-site symmetry $\hG$ \cite{ChenGuWen,FidKit,GuWen,BGK,WangLinGu,KTY2,WangGu2017,KapThorn17}. It is also given by an abelian group, where the group operation is stacking.

Every free fermionic system can be regarded as an interacting one (where the quartic and higher order interaction terms are set to zero), and this gives a homomorphism from the abelian group of free SRE phases to the abelian group of interacting ones (with the same symmetry). In general, this homomorphism is neither injective not surjective. The homomorphism may have a non-trivial kernel because some non-trivial free SRE phases can be destabilized by interactions. It may fail to be surjective because some interacting SRE phases are \emph{intrinsically interacting}, \emph{i.e.} cannot be realized by free fermions. The most familiar example of the former phenomenon occurs in 1d systems of class BDI \cite{FidKit}: while free SRE phases in this symmetry class are classified by $\ZZ$, the interacting ones are classified by $\ZZ_8$. An example of the latter phenomenon apparently occurs in dimension 6, where the cobordism classification of systems in class D predicts $\ZZ\times\ZZ$, while the free phases in the same symmetry class are classified by $\ZZ$. We study both phenomena more systematically in low dimensions. In particular, we will see that already in zero and one dimensions there exist fermionic SRE phases protected by a unitary symmetry which cannot be realized by free fermions.

To address such questions, it is very useful to have an efficient way to compute the interacting invariants of any given band Hamiltonian with any on-site symmetry $\hG$. One of the results of our paper is the computation of these invariants for arbitrary 0d and 1d band Hamiltonians. We also propose a partial answer in the 2d case. In the 1d case, we identify one of the invariants as a charge-pumping invariant.

The content of the paper as follows. In Section \ref{freesym}, we derive the classification of free SRE phases with a unitary symmetry $\hG$ in an arbitrary number of dimensions. In particular, we show that for $d=3$ all such phases are trivial. In Section \ref{interacting} we describe the map from free to interacting SRE phases for $d=0,1,$ and $2$.  Appendices \ref{pin} and \ref{charclass} contain some mathematical background. In Appendix \ref{betapump} we show that one of the invariants for free 1d SPT systems can be interpreted as a charge-pumping invariant.

A.\ T.\ is grateful to N. Strickland, M. Grant, and M. Wendt for their answers to the MathOverflow question Ref. \cite{strickland}. This research was supported in part by the U.S.\ Department of Energy, Office of Science, Office of High Energy Physics, under Award Number DE-SC0011632. The work of A.\ K.\  was partly performed at the Aspen Center for Physics, which is supported by National Science Foundation grant PHY-1607611. A.\ K.\ was also supported by the Simons Investigator Award.

\section{Free fermionic systems with a unitary symmetry}\label{freesym}
\subsection{Reduction to the ten-fold way}

In this section we show that the classification of free fermionic SRE systems with a unitary symmetry $\hG$ in dimension $d$ reduces to the classification of systems of class D, A, and C in the same dimension. The group $\hat G$ is assumed to be a compact Lie group. This includes finite groups as a special case. For simplicity, we show this for the case of 0d systems, from which the general case can be deduced. For systems of dimension $d>0$, the Majorana fermions have an additional index (the coordinate label). Accordingly, all matrices except $r(\hg)$ (defined below) become infinite. However, since the symmetry is on-site, all representation-theoretic manipulations remain valid, and the conclusions are unchanged.

Consider a general quadratic 0d Hamiltonian 
\begin{equation}\label{quadratic}
H=\frac{i}{2} A_{IJ}\Gamma^I \Gamma^J,
\end{equation}
where $A_{IJ}$, $I,J=1,\ldots,2N$ is a real skew-symmetric matrix and $\Gamma^I$ are Majorana fermions satisfying
\begin{equation}
\{\Gamma^I,\Gamma^J\}=2\delta^{IJ}.
\end{equation}
This Hamiltonian is known as the Majorana representation of the Bogoliubov-de Gennes Hamiltonian and may be straightforwardly obtained from its more familiar complex-fermion representation as, for example, in Ref. \cite{Ryuetal}. Suppose the Hamiltonian is invariant under a linear action of a group $\hG$:
\begin{equation}
\hg: \Gamma^I\mapsto \hR(\hg)^I_J\Gamma^J. 
\end{equation}
This defines a homomorphism $\hR:\hG\ra O(2N)$. 

Let us decompose $\hR$ into real irreducible representations of $\hG$. Suppose the irreducible representation $r_\alpha$ enters with multiplicity $n_\alpha$. The sum of all these copies of $r_\alpha$ will be called a block. It is clear that the Hamiltonian can only couple the fermions in the same block, so the matrix $A$ is block-diagonal. 

Let us focus on a particular block corresponding to an irreducible real  representation $r$.  There are three kinds of real irreducibles which are distinguished by the set of matrices which commute with all $r(\hg)$, $\hg\in \hG$ \cite{reptheory}. This set is known as the commutant of $r$. It is easy to see that it is closed under multiplication, and thus the commutant is an algebra. By Schur's lemma, if $r$ is irreducible, the commutant must be a real division algebra, so we have irreducibles of type $\RR$, $\CC$ and $\HH$, corresponding to the algebras of real numbers, complex numbers, and quaternions.\footnote{The reader may be more familiar with Schur's lemma for complex representations, where there is only one possible commutant: the unique complex division algebra $\CC$ corresponding to matrices proportional the the identity.} The corresponding block $A_r$ can be thought of as an operator on the space $r\otimes\RR^n$, where $n$ is the multiplicity of $r$. $\hG$-invariance of the Hamiltonian implies that this operator commutes with the $\hG$-action. The resulting constraint on $A_r$ depends on the type of the representation $r$.

If $r$ is of $\RR$-type, only scalar matrices commute with all $r(\hg)$. (Hence $r\otimes_\RR \CC$ is a complex irreducible representation of $\hG$. This is an equivalent characterization of $\RR$-type irreducibles.) Hence $A_r$ must have the form 
\begin{equation}
	A_r=1\otimes\cA,
\end{equation}
where $\cA$ is a real skew-symmetric matrix of size $n\times n$. There are no further constraints on $\cA$, so such a block can be thought of as describing $\dimr$ copies of a system of class D, i.e. a free fermion system whose only symmetry is the fermion parity. In particular,
\begin{equation}
    H=\frac{i}{2}A_{IJ}\Gamma^I\Gamma^J=\frac{i}{2}\sum_\mu^{\dim r}\cA_{ij}\Gamma^i_\mu\Gamma^j_\mu
\end{equation}
where we have relabeled fermions $\Gamma^I\mapsto\Gamma^i_\mu$ by indices $i=1,\ldots,n$ and $\mu=1,\ldots,\dim r$.

If $r$ is of $\CC$-type, then the algebra of matrices commuting with all $r(\hg)$ is spanned by $1$ and an element $J$ satisfying $J^2=-1$. Since $J^T$ must be proportional to $J$, this means that $J^T=-J$. The most general $\hG$-invariant $A_r$ must have the form
\begin{equation}
	A_r=1\otimes\cA+J\otimes\cC,
\end{equation}
where $\cA$ is skew-symmetric and $\cC$ is symmetric. We can equivalently parametrize such a Hamiltonian by a complex Hermitian matrix
\begin{equation}
    h=\cC+i\cA~.
\end{equation}
Upon complexification, we can decompose $r$ into eigenspaces of $J$ with eigenvalues $\pm i$. These eigenspaces are complex irreducible representations of $\hG$, and it is clear that they are conjugate to each other. We will denote them $q$ and ${\bar q}$. (An equivalent definition of a $\CC$-type representation is that $r\otimes_\RR\CC$ is a sum of two complex irreducible representations $q$ and ${\bar q}$ which are complex-conjugate and inequivalent). The $n\cdot \dimr$ Majorana fermions can be equivalently described by $\tfrac{1}{2}n\cdot \dimr$ complex fermions $\Psi^a_k$, $a=1,\ldots,n$, $k=1,\ldots, \frac12\dimr$ satisfying the commutation relations
\begin{equation}
\{\Psi^a_k,\bar\Psi^b_l\}=
\delta^a_b \delta^l_k.
\end{equation}
In terms of these fermions, the Hamiltonian takes the form
\begin{equation}
H=\sum_{k,a,b} \bar\Psi^b_kh_a^b\Psi^a_k.
\end{equation}
Thus a $\CC$-type block can be thought of as describing ${\rm dim}\, q=\frac12\dimr$ copies of a system of class A.

If $r$ is of $\HH$-type, then the algebra of matrices commuting with all $r(\hg)$ is spanned by $1$ and three elements $I,J,K$ which are skew-symmetric and obey the relations
\begin{equation}
I^2=J^2=K^2=-1,\quad IJ=K.
\end{equation}
Accordingly, $A_r$ must have the form
\begin{equation}
A_r=1\otimes\cA+I\otimes\cB+J\otimes\cC+K\otimes\cD,
\end{equation}
where $\cA$ is skew-symmetric and $\cB,\cC,\cD$ are symmetric. Equivalently, we can introduce a Hermitian $2n\times 2n$ matrix
\begin{equation}
Z=\begin{pmatrix} \cC+i\cA & \cB+i\cD \\ \cB-i\cD & -(\cC+i\cA)^T\end{pmatrix}. 
\end{equation}
This is the most general Hermitian matrix satisfying the particle-hole (PH) symmetry condition
\begin{equation}
C^\dagger Z^T C=-Z,
\end{equation}
where $C=i\sigma_2\otimes 1$. Since $C^*C=-1$, such a PH-symmetric system belong to class C.

To make this relationship with class C systems explicit, we again decompose $r\otimes_\RR\CC$ into a pair of complex-conjugate representations $q$ and ${\bar q}$. These two representations are equivalent, with the intertwiner being given by the tensor $I$. We also can think of $I$ as a non-degenerate skew-symmetric pairing $q\otimes q\ra \CC$. This implies that ${\rm dim}\, q$ is divisible by $2$ (and hence $\dimr$ is divisible by four).  As in the $\CC$-type case, we can describe the system by $n\cdot {\rm dim}\, q$ complex fermions. However, the presence of an $\hG$-invariant tensor $I$ means that the most general $\hG$-invariant Hamiltonian is
\begin{equation}
H=\bar\Psi (1\otimes h)\Psi+\frac12 \left(\Psi^T (I\otimes Y) \Psi+h.c.\right), 
\end{equation}
where $h$ is a Hermitian matrix, and $Y$ is a complex symmetric matrix. This is a BdG Hamiltonian, which can be re-written in terms of Dirac-Nambu fermions
\begin{equation}
\Phi=\begin{pmatrix} \Psi \\ (I\otimes 1)\bar\Psi^T\end{pmatrix}.
\end{equation}
The Dirac-Nambu spinors are defined so that the upper and lower components transform in the same way under $\hG$. They take values in $q\otimes\CC^2\otimes \CC^n$, where $\CC^2$ is the Dirac-Nambu space. The particle-hole (PH) symmetry acts by 
\begin{equation}
C: \Phi\mapsto \left(I\otimes\sigma_1\otimes 1\right) \bar\Phi^T
\end{equation}
and satisfies $C^2=-1$. In terms of Dirac-Nambu spinors, the Hamiltonian takes the form
\begin{equation}
H=\bar\Phi (1\otimes Z) \Phi, 
\end{equation}
where 
\begin{equation}
Z=\frac12 \begin{pmatrix} h & -Y^\dagger \\ -Y & -h^T\end{pmatrix}.
\end{equation}
Such matrices describe the most general class C system. Thus an $\HH$-type block can be thought of as describing ${\rm dim}\, q=\frac12\dimr$ copies of a system of class C.

\subsection{Classification of free SRE phases with a unitary symmetry}\label{freeclassif}

We always make the physically reasonable assumption that the generator of $\ZZ_2^F$ acts on all fermions by negation, i.e. 
\begin{equation}
\hR(P)=-1.
\end{equation}
The same must be true for all irreducible representations $r_\alpha$ which appear with nonzero multiplicity. We will call such irreducible representations \emph{allowed}. The set of all irreducible real representations of a compact group $\hG$ will be denoted $\Irr(\hG)$, while the set of all allowed irreducible real representations will be denoted $\Irr'(\hG)$. The set of allowed irreducible representations of type $K$ ($K=\RR,\CC,\HH$) will be denoted $\Irr'(\hG,K)$. If $\hG=\ZZ_2^F\times G$, we can identify $\Irr'(\hG,K)$ with the set $\Irr(G,K)$.

Let us recall the classification of class D, A, and C systems from the periodic table. Here we are listing only the ``strong'' invariants which do not depend on translational invariance.
\begin{center}
\begin{tabular}{c|cccccccc}
&$0$&$1$&$2$&$3$&$4$&$5$&$6$&$7$\\
\hline
Class D ($\RR$-type)&$\ZZ_2$&$\ZZ_2$&$\ZZ$&&&&$\ZZ$&\\
Class A ($\CC$-type)&$\ZZ$&&$\ZZ$&&$\ZZ$&&$\ZZ$&\\
Class C ($\HH$-type)&&&$\ZZ$&&$\ZZ_2$&$\ZZ_2$&$\ZZ$&
\end{tabular}
\end{center}
These results together with those of the previous subsection allow us to deduce the classification of free fermionic SREs with an arbitrary unitary symmetry $\hG$. In the physically interesting dimensions $d\leq 3$, the classification is given in Table \ref{table:G-classif}. This does not contradict the fact that there are interesting interacting fermionic 3d SREs.

\begingroup
\renewcommand{\arraystretch}{1.5}
\begin{table}
\begin{center}
\begin{tabular}{l|l}
$d=0$&$\oplus_{r\in \Irr'(\hG,\RR)} \ZZ_2\times \oplus_{r\in \Irr'(\hG,\CC)} \ZZ$\\
$d=1$&$\oplus_{r\in \Irr'(\hG,\RR)} \ZZ_2$\\
$d=2$&$\oplus_{r\in \Irr'(\hG)} \ZZ$\\
$d=3$&trivial
\end{tabular}
\end{center}
\caption{The classification of free phases protected by on-site unitary symmetry $\hat G$ in physical dimensions.}
\label{table:G-classif}
\end{table}
\endgroup

In what follows, an invariant attached to a particular irreducible representation $r_\alpha$ will be denoted $\varrho_\alpha$. Depending on the spatial dimension and the type of $r_\alpha$, $\varrho_\alpha$ will take values either in $\ZZ_2$ or $\ZZ$. An invariant of a free SRE phases will thus be a ``vector'' with components $\varrho_\alpha$. If $\hG$ is finite, then the number of allowed irreducible representations is finite, and the ``vector'' has a finite length. If $\hG$ is a compact Lie group, the number of allowed irreducible representations may be infinite, and then the space of ``vectors'' has infinite dimension (although all but a finite number of $\varrho_\alpha$ are zero for a particular SRE phase). These vectors can be interpreted as elements of the (twisted) equivariant K-theory, whose relevance to the classification of gapped band Hamiltonians is explained in \cite{FreedMoore}. 

The above results can be simplified a bit when $\hG$ is a product of $G$ and $\ZZ_2^F$. In this case the sums over allowed representations of $\hG$ can be replaced with the sums over all representations of $G$. 

The $\ZZ$ and $\ZZ_2$ invariants that appear in K-theory are relative invariants; that is, they detect something non-trivial about the junction between two phases. If one chooses a phase to regard as trivial (typically the phase containing the product state ground state in dimension $d>0$), the invariant for the junction of a phase $[H]$ with the trivial phase may be regarded as an absolute invariant of $[H]$.

\subsection{Examples}

Let us consider a few examples of free classifications for common symmetry groups.
\begin{itemize}
\item Superconductors with spin parity symmetry. $\hG=\ZZ_2^F\times\ZZ_2$. The action of $\ZZ_2^F$ on fermions is fixed, so we only need to choose the action of the second $\ZZ_2$. Overall, there are two allowed irreducible representations, both of them of $\RR$-type. Thus free phases with this symmetry are classified by $\ZZ_2\times\ZZ_2$ in 0d and 1d, and by $\ZZ\times\ZZ$ in 2d.
\item Charge-4e superconductors. $\hG=\ZZ_4$, where the $\ZZ_2$ subgroup is fermion parity. $\ZZ_4$ has three irreducible real representations, of dimensions $1$, $1$, and $2$, but only the $2$-dimensional representation is allowed. It is of $\CC$-type, hence free 0d and 2d phases with this symmetry are classified by $\ZZ$, while those in 1d have a trivial classification.
\item $\hG=\ZZ_2^F\times\ZZ_4$. Allowed irreducible representations of $\hG$ are equivalent to the $1$, $1$, and $2$ dimensional irreducible representations of $G=\ZZ_4$. Therefore the 0d classification is $\ZZ_2\times\ZZ_2\times\ZZ$, the 1d classification is $\ZZ_2\times\ZZ_2$, and the 2d classification is $\ZZ\times\ZZ\times\ZZ$.
\item Class A insulators. $\hG=U(1)$, with the obvious $\ZZ_2^F$ subgroup. There is one real representation for every non-negative integer, but only odd integers are allowed. All of these representations are of $\CC$-type, so free 0d phases with this symmetry are classified by $\ZZ^{\NN}$, that is, by a product of countably many copies of $\ZZ$. Note that although the symmetry is the same as for class A insulators, the classification is different. This is because it is usually assumed that complex fermions have charge $1$ with respect to $U(1)$, while we allow arbitrary odd charges. In 1d, there are no free phases with this symmetry, while in 2d there is again a $\ZZ^\NN$ classification.
\item $\hG=SU(2)$ with $\ZZ_2^F$ being the center. In this case, only representations of half-integer spin are allowed. All these representations are of $\HH$-type, hence all free Hamiltonians with this symmetry are in the same (trivial) phase in both 0d and 1d. In 2d, the classification is $\ZZ^\NN$.
\end{itemize}

\begingroup
\setlength{\tabcolsep}{0pt}
\renewcommand{\arraystretch}{1.1}
\begin{table}
\centering
\begin{tabular}{p{12mm}|p{5mm}p{20mm}p{20mm}p{20mm}p{20mm}p{20mm}p{35mm}}
$\hat G$&&$\ZZ_2^F$&$U(1)^F$&$SU(2)^F$&$\ZZ_2^F\times\ZZ_{2n}$&$\ZZ_{4n}^F$&$\ZZ_{4n+2}^F\simeq\ZZ_2^F\times\ZZ_{2n+1}$\\\\
$d=0$&&$\ZZ_2$&$\ZZ^\mathbb{N}$&$0$&$\ZZ_2^2\times\ZZ^{n-1}$&$\ZZ^n$&$\ZZ_2\times\ZZ^n$\\
$d=1$&&$\ZZ_2$&$0$&$0$&$\ZZ_2^2$&$0$&$\ZZ_2$\\
$d=2$&&$\ZZ$&$\ZZ^\mathbb{N}$&$\ZZ^\mathbb{N}$&$\ZZ^{n+1}$&$\ZZ^n$&$\ZZ^{n+1}$\\
$d=3$&&$0$&$0$&$0$&$0$&$0$&$0$
\end{tabular}
\caption{Free classification results for some common symmetry groups.}
\label{table:examples}
\end{table}
\endgroup

\section{Interacting invariants of band Hamiltonians}\label{interacting}

\subsection{Zero dimensions}

The only invariant of a general gapped fermionic 0d Hamiltonian with a unique ground state and symmetry $\hG$ is the charge\footnote{Recall that $H^1(\hG,U(1))$ is the group of one-dimensional unitary complex representations of $\hG$.} of the ground state 
\begin{equation}
\omega\in H^1(\hG,U(1)).
\end{equation}
As usual, this charge suffers from ambiguities, so it is better to consider the relative charge of two ground states. Let us compute this relative charge for the free Hamiltonian corresponding to a representation $\hR$. We decompose it into irreducibles, compute the charge in each sector separately, and then add up the results. 

Let us start with $\CC$-type representations. The corresponding Hamiltonian is described by a non-degenerate Hermitian matrix $h$ of size $n_r\times n_r$. Suppose we are given two such matrices $h$ and $h'$, with the number of negative eigenvalues $m_r$ and $m_r'$. We can consider a path deforming $h'$ to $h$. Every time an eigenvalue of $h'$ changes from a positive one to a negative one,  the ground state is multiplied by an operator
\begin{equation}
\prod_a\bar\Psi^a_i v^i,
\end{equation}
where $v^i$ is the corresponding eigenvector of $h$. Since $\bar\Psi^a_i$ transforms under $\hg\in\hG$ as
\begin{equation}
\bar\Psi^a_i\mapsto \bar q(\hg)^a_b \bar\Psi^b_i,
\end{equation}
the above operator has charge $\det\bar q(\hg)$. Thus a $\CC$-type irreducible  representation $r_\alpha$ contributes a relative charge
\begin{equation}
(\det\bar q_\alpha(\hg))^{\varrho_\alpha},
\end{equation}
where $\varrho_\alpha=m_\alpha-m'_\alpha\in\ZZ$ is the relative topological invariant of a pair of gapped class A Hamiltonians.

For an $\RR$-type representation $r$, the Hamiltonian is described by a non-degenerate skew-symmetric real matrix $A_{r,ij}$ of size $n_r\times n_r$. Any two such matrices $A_r$ and $A'_r$ are related by
\begin{equation}
A_r=O^T A'_r O,\quad O\in O(n_r). 
\end{equation}
To compute the relative charge of the ground states, we recall that the orthogonal group is generated by hyperplane reflections. Without loss of generality, we can assume that the hyperplane is orthogonal to the 1st coordinate axis. Let us compute the change in the ground state charge due to a reflection of the 1st coordinate axis. This corresponds to the following map on fermions:
\begin{equation}
\Gamma_a^1\mapsto -\Gamma_a^1,\quad a=1,\ldots,{\rm dim}\, r,
\end{equation}
while the rest of the fermions remain invariant. We need to treat separately the cases when ${\rm dim}\, r$ is even and when it is odd. 

If $\dimr$ is even, the map on fermions is in $SO(n_r\cdot \dimr)$, even though it arises from an element of $O(n_r)$ with determinant $-1$. On the Hilbert space, this map is represented by a bosonic operator proportional to 
\begin{equation}
\prod_{a=1}^{\dimr} \Gamma^1_a.
\end{equation}
This operator carries charge $\det\, r(\hg)$ under $\hg\in\hG$, hence the relative charge of the ground state corresponding to a hyperplane reflection is $\det\, r(\hg)$.  

If $\dimr$ is odd, the map on fermions is an orthogonal transformation with determinant $-1$, and thus must be represented on the Hilbert space by a fermionic operator. This fermionic operator is proportional to
\begin{equation}
\prod_{j=2}^{n_r}\prod_{a=1}^\dimr \Gamma^j_a .
\end{equation}
It carries charge $(\det\, r(\hg))^{n_r-1}=\det\, r(\hg)$ under $\hg\in\hG$. Hence the relative charge of the ground state is again $\det\, r(\hg)$. 

We conclude that when $O\in O(n_r)$ is a hyperplane reflection, the relative charge of the ground state under $\hg\in\hG$ is $\det\, r(\hg)$. Since $\det\,  r(\hg)=\pm 1$ and every element of $SO(n_r)$ is a product of an even number of hyperplane reflections, this implies that the relative charge is trivial when $O\in SO(n_r)$. Since every element of $O(n_r)$ is a product of a hyperplane reflection and an element of $SO(n_r)$, the relative charge of the ground state for an $O$ which is not in $SO(n_r)$ is $\det\, r(\hg)$. 

To summarize, the relative charge contribution from an $\RR$-type representation $r_\alpha$ is
\begin{equation}
\left(\det\, r_\alpha(\hg)\right)^{\varrho_\alpha},
\end{equation}
where $\varrho_\alpha\in\ZZ_2$ is the relative invariant of a pair of gapped class D Hamiltonians.

Finally, $\HH$-type representations do not contribute to the relative charge since all 0d class C systems are deformable into each other.

In summary, the map from free to interacting phases in 0d is
\begin{equation}\label{map0d}
\{\varrho_\alpha\}\mapsto \omega(\hg)=\prod_{\alpha\in Irr'(\hG,\RR)}(\det r_\alpha(\hg))^{\varrho_\alpha}\prod_{\alpha\in Irr'(\hG,\CC)}(\det \bar q_\alpha(\hg))^{\varrho_\alpha}.
\end{equation}

In what follows, we often find it more convenient to identify $U(1)$ with $\RR/\ZZ$, i.e. write the abelian group operation on 1-cocycles additively rather than multiplicatively. This amounts to taking the logarithm of both sides of (\ref{map0d}) and dividing by $2\pi i$. Then $\omega$ becomes as sum of two terms, $\omega=\omega_1+\omega_2$. The first term 
\begin{equation}
\omega_1(\hg)=\sum_{\alpha\in\Irr'(\hG,\RR)}\frac{1}{2\pi i} \varrho_\alpha\log\det r_\alpha(\hg)
\end{equation}
can be interpreted as the weighted sum of the 1st Stiefel-Whitney classes of the representations $r_\alpha$ (see Appendix \ref{charclass} for an explanation of this terminology). More precisely, the 1st Stiefel-Whitney class $w_1(r_\alpha)$ is an element of $H^1(\hG,\ZZ_2)$, while $\omega_1$ involves the corresponding class in $H^1(\hG,\RR/\ZZ)$ which we denote $w_1^{U(1)}(r_\alpha)$:
\begin{equation}
\omega_1=\sum_{\alpha\in\Irr'(\hG,\RR)} \varrho_\alpha w_1^{U(1)}(r_\alpha)\in H^1(\hG,\RR/\ZZ).
\end{equation}
The 2nd term which arises from $\CC$-type representations can be interpreted in terms of the 1st Chern class of the complex representations $q_\alpha$:
\begin{equation}
\omega_2=\sum_{\alpha\in\Irr'(\hG,\CC)}  \beta^{-1}(c_1(\varrho_\alpha q_\alpha))\in H^1(\hG,\RR/\ZZ).
\end{equation}
Here $\beta^{-1}$ is the inverse of the Bockstein isomorphism $\beta: H^1(\hG,\RR/\ZZ)\ra H^2(\hG,\ZZ)$. In the 0d case, it seems superfluous to express determinants in terms of Stiefel-Whitney and Chern classes, but in higher dimensions characteristic classes of representations become indispensable. They are briefly reviewed in Appendix \ref{charclass}. 

It is clear that the map from $\{\varrho_\alpha\}$ to $\omega$ is many-to-one for almost all $\hG$. In fact, for Lie group symmetries, such as $U(1)$ or $SU(2)$, a single interacting phase corresponds to an infinite number of free phases.

More surprisingly, the map may fail to be surjective. A class $\omega\in H^1(\hG,U(1))$ defines a one-dimensional complex representation $q$ of $\hG$. If this representation is allowed (i.e. if $\omega(P)=-1$), we can take a complex fermion $\bar\Psi$ and its Hermitian conjugate $\Psi$  and let them transform in the representations $q$ and $\bar q$, respectively. Now the two $\hG$-invariant Hamiltonians
\begin{equation}
H_\pm=\pm \left(\bar\Psi\Psi-\frac12\right)
\end{equation}
have relative ground-state charge $\omega$. But if the representation $q$ is not allowed, $\omega(P)=1$, then the situation is more complicated. For certain $\hG$ there are no allowed one-dimensional representations at all, but one could try to use higher-dimensional allowed representation to get the relative ground-state charge $\omega$.

Let us exhibit an example of a group $\hG$  where certain relative charges $\omega$ cannot be obtained from free systems. This shows that the map from free to interacting 0d phases is not surjective, in general. Consider extending the group $G=\ZZ_4\times \ZZ_4$ by $\ZZ_2$. If the extension class in $H^2(G,\ZZ_2)$ maps to a non-trivial element in $H^2(G,U(1))$, the group $\hG$ may be presented in terms of generators $A,B,P$, where $P$ is central and
\begin{equation}
	P^2=A^4=B^4=1\qquad\text{and}\qquad AB=PBA.
\end{equation}
The group of one-dimensional representations of $\hG$ is then the same as the group of one-dimensional representations of $G$, i.e. $\ZZ_4\times\ZZ_4$, defined by $q(A),q(B)\in\{\pm 1,\pm i\}$. All sixteen  of these are disallowed, as $q(P)=+1$. Up to equivalence, only four irreducible representations remain. They are two-dimensional and of the form $q(P)=-\mathds{1}_2$ (allowed), $q(A)=i^a\sigma_z$, and $q(B)=i^b\sigma_x$, for $a,b\in\{0,1\}$. Each is related to a complexification of a real irreducible representation $r$ by $r_\CC=q\oplus\bar q$ and has $\det q(\hg)\in\{\pm 1\}$. This means that twelve out of sixteen cocycles (those with $\omega(\hg)=\pm i$ for some $\hg$) do not arise from free systems.

\subsection{One dimension}\label{1d}

Let us begin by recalling invariants of interacting fermionic SRE phases in 1d and their interpretation in terms of boundary zero modes. 
Any fermionic 1d SRE phase has an invariant $\gamma\in\ZZ_2$ \cite{FidKit}.
(From now on, we will write $\ZZ_2$ additively, i.e. identify it with the set $\{0,1\}$, unless stated otherwise.) It tells us whether the number of fermionic zero modes on the boundary is even or odd. Algebraically, if $\gamma=0$, the algebra of boundary zero modes $A_b$ is a matrix algebra, while for $\gamma=1$ it is a sum of two matrix algebras. In both cases $A_b$ is simple provided we regard it as a $\ZZ_2^F$-graded algebra. In the case $\gamma=0$, the graded center of $A_b$ is isomorphic to $\CC$, while for $\gamma=1$ it is isomorphic to $\Cl(1)$. The odd generator of $\Cl(1)$ is denoted $\hat Z$.

If the system also has a unitary symmetry $\hG$, then there are further invariants whose form depends on the value of $\gamma$ \cite{FidKit}. If $\gamma=0$, the additional  invariant is $\halpha\in H^2(\hG,U(1))$. If $\gamma=1$, the additional invariants are a homomorphism $\mu:\hG\ra \ZZ_2$ such that $\mu(P)=1$ (the generator of $\ZZ_2$) and $\alpha\in H^2(G,U(1))$. A homomorphism $\mu$ allows one to define an isomorphism $\hG\simeq G\times\ZZ_2^F$ as follows:
\begin{equation}
\hg\mapsto (g,\mu(\hg)).
\end{equation}
So if $\hG$ is not isomorphic to the product $G\times\ZZ_2^F$, the case $\gamma=1$ is impossible. 

Note that there is a homomorphism $H^2(\hG,U(1))\ra H^1(G,\ZZ_2)$ whose kernel is non-canonically isomorphic to $H^2(G,U(1))$. To see this, let us define the group law on $\hG$ using a $\ZZ_2$-valued 2-cocycle $\rho$ on $G$:
\begin{equation}
(g,\eps)\circ (g',\eps')=(gg',\eps+\eps'+\rho(g,g')),\quad g,g'\in G,\quad \eps,\eps'\in \{0,1\}.
\end{equation}
Then $\halpha$ can be parameterized by a pair of cochains $(\alpha,\beta)\in C^2(G,U(1))\times C^1(G,\ZZ_2)$ satisfying $\delta\beta=0$ and $\delta\alpha=\frac12\rho\cup\beta$, modulo $\alpha\mapsto\alpha+\delta\lambda$, $\lambda\in C^1(G,U(1))$. The map from $H^2(\hG,U(1))$ to $H^1(G,\ZZ_2)$ sends the pair $(\alpha,\beta)$ to $\beta$. 

The boundary interpretation of the additional invariants also depends on whether $\gamma=0$ or $\gamma=1$. For $\gamma=0$, the algebra $A_b$ is a matrix algebra, and therefore $\hG$ acts on it by conjugation:
\begin{equation}
\hg: a\mapsto V(\hg) a V(\hg)^{-1},\quad a\in A_b.
\end{equation}
One can even choose the invertible elements $V(\hg)\in A_b$ to be unitary ($A_b$ is actually a $C^*$-algebra, so the notion of a unitary element makes sense).
The elements $V(\hg)$ are well-defined up to a $U(1)$ factor and satisfy
\begin{equation}
V(\hg)V(\hg')=\halpha(\hg,\hg') V(\hg \hg'),
\end{equation}
where $\halpha$ is a 2-cocycle on $\hG$.

On the other hand, if $\gamma=1$, then the same considerations apply to the even part of the graded algebra $A_b$, and one gets an invariant $\alpha\in H^2(G,U(1))$ in the same way. In addition, one can ask how the group $\hG$ acts on the odd central element $\hat Z\in A_b$. One must have
\begin{equation}
\hg: \hat Z\mapsto (-1)^{\mu(\hg)}\hat Z,
\end{equation}
where $\mu:\hG\ra\ZZ_2$ is a homomorphism satisfying $\mu(P)=1$. 

As explained above, free SRE 1d systems with symmetry $\hG$ are classified by a sequence of invariants $\varrho_\alpha\in\ZZ_2$, one for each real irreducible representation of $\hG$ of $\RR$-type. The physical meaning of $\varrho_\alpha$ is simple. The group $\hG$ acts on the boundary zero modes (assumed to form a Clifford algebra) via a real representation\footnote{One should not confuse the ``boundary'' representation $\cR$ with the on-site representation $\hR$. The former can be odd-dimensional, while the latter is always even-dimensional. Also, $\hR$ takes values in $SO(2N)$, while $\cR$ in general takes values in the orthogonal group.}
\begin{equation}\label{boundaryR}
\cR=\oplus \nu_\alpha r_\alpha.
\end{equation}
The integer $\nu_\alpha$ reduced modulo $2$ is the free topological invariant $\varrho_\alpha$ discussed in Section \ref{freeclassif}.

Let us now describe the map from free to interacting invariants. For a free system, the algebra of boundary zero modes is $A_b=\Cl(M)$, so one has $\gamma=M\, {\rm mod}\, 2$. Equivalently, using the decomposition (\ref{boundaryR}), we get
\begin{equation}
\gamma=\sum_\alpha \varrho_\alpha\,  \dimr_\alpha\, \mod 2.
\end{equation}

Now let us determine the remaining invariants.
Consider the case $\gamma=0$ first. Then $O(M)$ is a non-trivial extension of $SO(M)$ by $\ZZ_2$. We can interpret $A_b=\Cl(M)$ as the algebra of operators on a Fock space of dimension $2^{M/2}$, and the group $\hG$ acts projectively on this space. The cohomology class of the corresponding cocycle is $\halpha$. Clearly, it is completely determined by the representation $\cR:\hG\ra O(M)$. 

From the group-theoretic viewpoint, a projective action of $\hG$ on the Fock space is the same as a homomorphism $\hG\ra Pin_c(M)$, where $Pin_c(M)$ is a certain non-trivial extension of $O(M)$ by $U(1)$. $Pin_c(M)$ and related groups are reviewed  in Appendix \ref{pin}. Thus $\halpha$ is the obstruction to lifting $\cR$ to a homomorphism $\hG\ra Pin_c(M)$. As discussed in Appendix \ref{charclass}, this obstruction is the image of the 2nd Stiefel-Whitney class of $\cR$ under the homomorphism $H^2(\hG,\ZZ_2)\ra H^2(\hG,U(1))$. We denote it $w_2^{U(1)}(\cR)$. The Whitney formula for Stiefel-Whitney classes says
\begin{equation}\label{w2}
w_2(\cR)=w_2\left(\oplus \nu_\alpha r_\alpha\right)=
\sum_\alpha \varrho_\alpha w_2(r_\alpha)+\sum_{\alpha<\beta}  \varrho_\alpha\varrho_\beta w_1(r_\alpha)\cup w_1(r_\beta).
\end{equation}
Therefore
\begin{equation}\label{halpha}
\halpha=w^{U(1)}_2(\cR)=\sum_\alpha \varrho_\alpha w_2^{U(1)}(r_\alpha)+\sum_{\alpha<\beta}  \frac12 \varrho_\alpha\varrho_\beta w_1(r_\alpha)\cup w_1(r_\beta).
\end{equation}
Note that $Pin_c(M)$ is a $\ZZ_2$-graded group, i.e. it is equipped with a homomorphism to $\ZZ_2$. The value of this homomorphism tells us if $V(\hg)$ is an even or odd element in $\Cl(M)$. 
It is easy to see that this homomorphism is precisely $\beta(g)$. On the other hand, as explained in Appendix \ref{charclass}, the said homomorphism is simply 
$\det \cR(\hg)$. Thus
\begin{equation}
\beta=w_1(\cR)=\sum_\alpha\varrho_\alpha w_1(r_\alpha).
\end{equation}
In Appendix \ref{betapump} we give an alternate characterization of $\beta$ as a charge-pumping invariant.

Now consider the case $\gamma=1$, where $A_b\simeq \Cl(M)$ with odd $M$. In agreement with \cite{FidKit}, the map $\hg\mapsto\det\cR(\hg)$ defines a splitting of $\hG$, i.e. an isomorphism $G\times\ZZ_2^F\simeq\hG$. This means 
\begin{equation}
\mu=w_1(\cR)=\sum_\alpha \varrho_\alpha w_1(r_\alpha).
\end{equation}
We can define a new representation $\tilde\cR:G\ra SO(M)$ by
\begin{equation}
\tilde\cR(g)=\cR(\hg) \det\cR(\hg).
\end{equation}
Here $\hg\in\hG$ is any lift of $g\in G$. Thus we get a homomorphism
\begin{equation}
G\times\ZZ_2^F\ra O(M)\simeq SO(M)\times\ZZ_2^F,\quad (g,\eps)\mapsto (\tilde\cR(g),\eps).
\end{equation}
By definition, $\alpha$ is the obstruction for lifting $\tilde\cR$ to a homomorphism $G\ra Spin_c(M)$. Thus
\begin{equation}
\alpha=w_2^{U(1)}(\tilde\cR).
\end{equation}
Using a formula for Stiefel-Whitney classes of a tensor product (see Appendix \ref{charclass}), one can show that $w_2(\tilde\cR)=w_2(\cR)$, and thus one can also write
\begin{equation}
\alpha=w_2^{U(1)}(\cR). 
\end{equation}

We note that the map from free to  interacting 1d SRE phases is compatible with the stacking law derived in \cite{KTY2,GK15}. For example, if we consider for simplicity the case $\gamma=0$, then the stacking law takes the form
\begin{equation}\label{stackinglaw1d}
\halpha\circ\halpha'=\halpha+\halpha'+\tfrac{1}{2} \beta\cup\beta' 
\end{equation}
On the other hand, stacking two SRE systems characterized by representations $\cR$ and $\cR'$ gives an SRE system corresponding to the representation $\cR\oplus\cR'$.
If we set $\alpha=w_2^{U(1)}(\cR)=\frac12 w_2(\cR)$ and $\beta=w_1(\cR)$, then the stacking law (\ref{stackinglaw1d}) follows from the Whitney formulas 
\begin{eqnarray}
&w_1(\cR\oplus\cR') =w_1(\cR)+w_1(\cR'),\\
&w_2(\cR\oplus\cR')=w_2(\cR)+w_2(\cR')+w_1(\cR)\cup w_1(\cR').
\end{eqnarray}

It is clear that the map from free to interacting phases is not injective. Let us discuss surjectivity. We have seen that for free systems the invariants $\halpha$ and $\alpha$ are always of order $2$. Hence to get an example of a fermionic SRE phase which cannot be realized by free fermions, it is sufficient to pick a $\hG$ and a non-trivial 2-cocycle which is not of order $2$. For example, if we take $\hG=\ZZ_2^F\times\ZZ_3\times\ZZ_3$, and take $\alpha$ to be any non-trivial element of $H^2(\ZZ_3\times\ZZ_3,U(1))=\ZZ_3$, then such a phase cannot be realized by free fermions. 

One might hope that perhaps every $\halpha$ or $\alpha$ of order $2$ can be realized by free fermions, but this is not the case either. The reason for this is that for any orthogonal representation $\cR$ of $\hG$, the 2-cocycle $w_2(\cR)$ satisfies some relations \cite{strickland}. This is explained in Appendix \ref{charclass}. These relations need not hold for a general 2-cocycle on $\hG$.  Unfortunately, the simplest example of $\hG$ for which this happens is rather non-trivial \cite{metacyclic}.

While not every fermionic 1d SRE phase can be realized by free fermions, every fermionic 1d SRE phase with $\hG\simeq G\times\ZZ_2^F$ can be realized by stacking bosonic 1d SRE phases with free fermions. First, we can change $\gamma$ of an SRE phase at will by stacking with the Kitaev chain. If we make $\gamma=0$ by such stacking, then we can change $\beta$ at will by stacking with two copies of the Kitaev chain on which the group $G$ acts by 
\begin{equation}
(\gamma_1,\gamma_2)\mapsto ((-1)^{\beta(g)}\gamma_1,\gamma_2).
\end{equation} 
Finally, since $\alpha$ is an arbitrary element of $H^2(G,U(1))$ in this case, one can change it at will by stacking with bosonic SRE phases with symmetry $G$.

\subsection{Two dimensions}\label{2d}

To every fermionic 2d SRE phase one can attach an integer invariant $\kappa$. It measures the chiral central charge for the boundary CFT.

If the SRE has a unitary symmetry $\hG$, there are further invariants. For simplicity, let us assume that we are given an isomorphism $\hG\simeq G\times\ZZ_2^F$. We will also assume that $G$ is finite, rather than merely compact. Then the invariants are a 1-cocycle $\gamma\in H^1(G,\ZZ_2)$, a 2-cocycle $\beta\in H^2(G,\ZZ_2)$, and a 3-cochain $\alpha\in C^3(G,U(1))$ satisfying 
\begin{equation}
\delta\alpha=\frac{1}{2}\beta\cup\beta.
\end{equation}
There are certain non-trivial identifications on these data, see \cite{GuWen,GK15}. The abelian group structure corresponding to stacking the systems is also quite non-trivial. We just note for future use that if we ignore $\alpha$, the group law is
\begin{equation}
(\beta,\gamma)+(\beta',\gamma')=(\beta+\beta'+\gamma\cup\gamma',\gamma+\gamma').
\end{equation}
The physical meaning of these invariants is somewhat complicated, with the exception of $\gamma(g)$: it measures the number of Majorana zero modes on a $g$-vortex, reduced modulo $2$. 

On the other hand, a free 2d SRE is characterized by a sequence of invariants $\varrho_\alpha\in\ZZ$, one for each real irreducible representation of $G$. 

It is easy to determine the chiral central charge $\kappa$ for such a free SRE. A basic system of class D  has $\kappa=1/2$. For example, a $p+ip$ superconductor has a single chiral Majorana fermion on the boundary which has chiral central charge $1/2$.\footnote{In the literature on fermionic SRE phases, it is common to re-write systems of class D, which only have a $\ZZ_2^F$ symmetry, as systems with both a $U(1)$ symmetry and a particle-hole symmetry \cite{Bernevig,Ryuetal}. This entails doubling the number of degrees of freedom, and therefore doubling $\kappa$.} A basic system of class A has $\kappa=1$. For example, the basic Chern insulator has a single chiral complex fermion on the boundary which has chiral central charge $1$. Two basic class C  systems\footnote{Since ${\rm dim}\, q$ is even for $\HH$-type representations, only an even number of class C systems can occur.} have chiral central charge $2$. For example, two copies of the basic Chern insulator can be regarded as the basic class C system tensored with a two-dimensional representation of $SU(2)$, and thus has $\kappa=2$. Consequently, the chiral central charge is given by
\begin{equation}
\kappa=\frac12\sum_{r_\alpha\in \Irr(G)} \varrho_\alpha {\rm dim}\, r_\alpha 
\end{equation}

The other interacting invariants are harder to deduce. We will propose natural candidates for $\gamma$ and $\beta$ based on experience with lower-dimensional cases.

Given an orthogonal representation $r:G\ra O(n)$ we can define a 1-cocycle
\begin{equation}
\det r(g)\in H^1(G,\ZZ_2).
\end{equation}
It is sometimes called the 1st Stiefel-Whitney class of $r$, for reasons explained in Appendix \ref{charclass}. We will denote it $w_1(r)$. For irreducible representations of type $\CC$ and $\HH$ it is trivial.\footnote{For $\CC$-type representations, we have $\det r(g)=\det q(g)\det {\bar q}(g)=1$, while for $\HH$-type representations $\det q(g)=1$ since $q(g)$ takes values in the unitary symplectic group.}  

Similarly, we can define the 2nd Stiefel-Whitney class of $G$ as an obstruction to lifting $r:G\ra O(n)$ to $\tilde r_+: G\ra Pin_+(n)$. One can lift each $r(g)$ to an element $\tilde r_+(g)\in Pin_+$, but the composition law will only hold up to a 2-cocycle $\lambda(g,g')$ with values in $\pm 1$. Thus we get a well-defined element $w_2(r)\in H^2(G,\ZZ_2)$. One might also consider an  obstruction to lifting $r$ to a homomorphism $\tilde r_-: G\ra Pin_-(n)$, but it is expressed in terms of $w_2(r)$ and $w_1(r)$ (namely, the $Pin_-$ obstruction is $w_2+w_1^2$). 
 
A natural guess for the contribution of an irreducible $r_\alpha$ to $\gamma$ is $\varrho_\alpha w_1(r_\alpha)$. Assuming this, the formula for the invariant $\gamma$ is
\begin{equation}
\gamma=\sum_{r_\alpha\in\Irr(G,\RR)} \varrho_\alpha w_1( r_\alpha)=w_1(\cR),
\end{equation}
where we defined a ``virtual representation''\footnote{The word ``virtual'' reflects the fact that the numbers $\varrho_\alpha$ can be both positive and negative. Thus $\cR$ is best thought of as an element of the K-theory of the representation ring of $G$.}
\begin{equation}
\cR=\oplus_\alpha\varrho_\alpha r_\alpha.
\end{equation}
Note that only $\RR$-type representations contribute to $\gamma$, since only those representations can have nonzero $w_1(r)$. On the other hand, $\cR$ includes all representations.

There are two natural guesses for the contribution of a single irreducible $r$ to $\beta$: $w_2(r)$ or $\tilde w_2(r)=w_2(r)+w_1(r)^2$. To derive $\beta$ for a general virtual representation $\cR$, we note that  the Whitney formula for Stiefel-Whitney classes says 
\begin{equation}
w_2(\cR+\cR')=w_2(\cR)+w_2(\cR')+w_1(\cR)\cup w_1(\cR').
\end{equation}
The same formula applies to $\tilde w_2(r)$. This formula looks just like the stacking law for $\beta$ and $\gamma$, if we identify $\gamma$ with $w_1$ and $\beta$ with $w_2$ (or $\tilde w_2$).
Hence for a general $\cR$ we have either $\beta(\cR)=w_2(\cR)$ or $\beta(\cR)=w_2(\cR)+w_1(\cR)^2$.

A non-trivial check on both of these candidates is that they are compatible with the group supercohomology equation. This equation implies that $\beta\cup\beta\in H^4(G,\ZZ_2)$ maps to a trivial class in $H^4(G,U(1))$. This is automatically satisfied for both $\beta=w_2(\cR)$ and $\beta=w_2(\cR)+w_1(\cR)^2$, as shown in Appendix \ref{charclass}.
%

Is there any way to decide between the two candidates for $\beta$? Not without understanding better the physical meaning of $\beta$. Indeed, formally, a change of variables $\beta\mapsto \beta+\gamma\cup\gamma$ is an automorphism of the group of fermionic SRE phases in 2d. This automorphism maps one candidate for $\beta$ to the other one. Thus formally they are equally good. One can pick one over another only if one assigns $\beta$ a particular physical meaning. The same is even more true about $\alpha\in C^3(G,U(1))$, since it depends on various choices in a complicated way. 

Let us make a few remarks about surjectivity of the map from free to interacting SRE phases in the 2d case. It is clear that every value of the parameter $\gamma\in H^1(G,\ZZ_2)$ can be realized by free fermionic systems. One can just take two copies of the basic system of class A with opposite values of the chiral central charge $\kappa$ (for example, a $p+ip$ superconductor stacked with a $p-ip$ superconductor) and let $G$ act only on the first copy via a 1-dimensional real representation of $G$ given by the 1-cocycle $\gamma$. This construction was used in Ref. \cite{LevinGu} for the case $G=\ZZ_2$. 

One can also ask if every $\beta$ that  solves the supercohomology equation can be realized by free fermions. The answer appears to be no \cite{strickland}, for a sufficiently complicated $G$. The reason is again some highly non-trivial relations satisfied by Stiefel-Whitney classes. Thus not all supercohomology phases in 2d can be realized by free fermions. At the moment we do not know how to find a concrete example of a finite group $G$ for which this happens. It would be interesting to study this question further and in particular determine both $\alpha$ and $\beta$ for a general 2d band Hamiltonian with symmetry $G$.

\appendix

\section{Pin groups}\label{pin}

Here we review the definition and some properties of $Pin$ groups following Ref.  \cite{ABS}.
Just as $Spin(M)$ is a non-trivial extension of $SO(M)$ by $\ZZ_2$, $Pin_+(M)$ and $Pin_-(M)$ are extensions of $O(M)$ by $\ZZ_2$. Since $O(M)$ has two connected components, so do $Pin_\pm (M)$. The connected component of the identity for both $Pin_+(M)$ and $Pin_-(M)$ is $Spin(M)$. 

The groups $Pin_\pm (M)$ can be defined using the Clifford algebra $\Cl(M)$. To define $Pin_+(M)$, one considers the Clifford algebra for the positive metric:
\begin{equation}
\{\Gamma^I,\Gamma^J\}=2\delta^{IJ},\quad I,J=1,\ldots,M.
\end{equation}
This is a $\ZZ_2$-graded algebra. For any $a\in Cl(M)$ we let $\eps(a)=a$ if $a$ is even and $\eps(a)=-a$ if $a$ is odd.
Invertible elements in the Clifford algebra form a group. $Pin_+(M)$ is a subgroup generated by elements of the form $\vslash=\Gamma^I v^I$, where $v_I$ is a unit vector in $\RR^M$. To define the homomorphism $Pin_+(M)\ra O(M)$, consider the ``twisted conjugation map''
\begin{equation}
\Gamma^J\mapsto \eps(a) \Gamma^J a^{-1},\quad a\in Cl(M).
\end{equation} 
If $a=\vslash$, then this map becomes
\begin{equation}
\Gamma^J \mapsto -\vslash \Gamma^J \vslash^{-1} =\Gamma^J-2v^J \vslash .
\end{equation}
This is a hyperplane reflection on the space spanned by $\Gamma^J$. Since the  whole group $O(M)$ is generated by hyperplane reflections, twisted conjugation by elements of $Pin_+(M)$ gives a surjective homomorphism from $Pin_+(M)$ to $O(M)$. The kernel of this map is the $\ZZ_2$ generated by $-1$. 
The subgroup $Spin(M)\subset Pin_+(M)$ consists of products of an even number of hyperplane reflections. Note that every hyperplane reflection $\vslash$ squares to the identity in $Pin_+(M)$. 

The group $Pin_-(M)$ is defined similarly, except that one starts with the ``negative'' Clifford algebra
\begin{equation}
\{\Gamma^I,\Gamma^J\}=-2\delta^{IJ},\quad I,J=1,\ldots,M.
\end{equation}
In this case, hyperplane reflections $\vslash$ square to $-1$, which generates the kernel of the homomorphism $Spin(M)\ra SO(M)$. In other words, for $Pin_-(M)$, hyperplane reflections square to fermion parity. 

Finally, the group $Pin_c(M)$ is defined as $(Pin_+(M)\times U(1))/\ZZ_2^{diag}$, and its subgroup $Spin_c(M)\subset Pin_c(M)$ is defined as $(Spin(M)\times U(1))/\ZZ_2$. $Pin_c(M)$ is an extension of $O(M)$ by $U(1)$, while $Spin_c(M)$ is an extension of $SO(M)$ by $U(1)$. It is easy to show that the group $(Pin_-(M)\times U(1))/\ZZ_2$ is isomorphic to $Pin_c(M)$. The significance of $Pin_c(M)$ is the following: if we regard the complexification of the Clifford algebra as the algebra of observables of a fermionic system, then $Pin_c(M)$ can be identified with the subgroup of those unitaries which act linearly on the generators of the Clifford algebra. Thus lifting a real linear action of a group $G$ on the Clifford generators $\Gamma^I$ to a unitary action on the Fock  space is equivalent to lifting the corresponding homomorphism $G\ra O(M)$ to a homomorphism $G\ra Pin_c(M)$. Similarly, if we are given a homomorphism $G\ra SO(M)$, lifting it to a unitary action on the Fock space is the same as lifting it to a homomorphism $G\ra Spin_c(M)$.

\section{Characteristic classes of representations of finite groups}\label{charclass}

The theory of characteristic classes of vector bundles (a classic reference is \cite{MilnorStasheff}) is familiar to physicists. A version of this construction also gives rise to characteristic classes of representations of a finite group which take values in cohomology of the said group \cite{charclassesofreps}. Real representations give rise to Stiefel-Whitney and Pontryagin classes, while complex representations give rise to Chern classes.

To define these classes, it is best to think of a real representation of $G$ of dimension $n$ as a homomorphism $R: G\ra O(n)$, which then induces a continuous map of classifying spaces $\tilde R: BG\ra BO(n)$. The map $\tilde R$ is defined up to homotopy only, but this suffices to define cohomology classes on $BG$ by pull-back from $BO(n)$. Any cohomology class $\omega$ on $BO(n)$ thus defines a cohomology class $\tilde R^*\omega$ on $BG$. Cohomology classes on $BO(n)$ are precisely characteristic classes of real vector bundles, and their pull-backs via $\tilde R$ are called characteristic classes of the representation $R$. Similarly, given a complex representation $R:G\ra U(n)$, we get a continuous map $\tilde R:BG\ra BU(n)$, and can define Chern classes of $R$ by pull-back. 

In low dimensions, these classes have a concrete representation-theoretic interpretation. For example, the 1st Stiefel-Whitney class $w_1(R)\in H^1(G,\ZZ_2)$ of a real representation $R$ is the obstruction for $R:G\ra O(n)$ to descend to homomorphism $R':G\ra SO(n)$. Obviously $w_1(r)(g)$ is given by $\det R(g)$. 

Similarly, the 1st Chern class $c_1(R)\in H^2(G,\ZZ)$ of a complex representation $R$ can be interpreted as an obstruction for  $R$ to descend to $R':G\ra SU(n)$. The  obstruction $\det\, R(g)$ is a 1-cocycle on $G$ with values in $U(1)$. The corresponding class in $H^2(G,\ZZ)$ is obtained by applying the Bockstein homomorphism (which for finite groups is an isomorphism). Explicitly:
\begin{equation}
c_1(R)(g,h)=\frac{1}{2\pi i}\left(\log\det R(gh)-\log\det R(g)-\log\det R(h)\right). 
\end{equation}

The 2nd Stiefel-Whitney class $w_2(R)\in H^2(G,\ZZ_2)$ is an obstruction to lifting $R$ to a homomorphism $R': G\ra Pin_+(n)$. One can always define $R'$ as a projective representation, and the corresponding 2-cocycle represents $w_2(R)$. The image of $w_2(R)$ in $H^2(G,U(1))$ under the embedding $\ZZ_2\ra U(1)$ is an obstruction to lifting $R$ to a homomorphism $R':G\ra Pin_c(n)$. In the main text, it is denoted $w_2^{U(1)}(R)$. By the isomorphism $H^2(G,U(1))\simeq H^3(G,\ZZ)$ (valid for finite groups), this class can be interpreted as an element of $H^3(G,\ZZ)$. Then it is known as the 3rd integral Stiefel-Whitney class $W_3$. 

By functoriality, known relations between cohomology classes of $BO(n)$ and $BU(n)$ lead to relations between characteristic classes of representations. Let us describe those of them which we have used in the main text. First of all, the Whitney formula expresses Stiefel-Whitney (or Chern) classes of $R+R'$ in terms of Stiefel-Whitney (or Chern) classes of $R$ and $R'$:
\begin{equation}
w_k(R+R')=\sum_{p=0}^k w_p(R)\cup w_{k-p}(R').
\end{equation}
There are also more complicated formulas expressing characteristic classes of $R\otimes R'$ in terms of those of $R$ and $R'$ \cite{MilnorStasheff}. We will only need a particular case: let $R$ be a real representation of odd dimension $M$, and $L$ be a one-dimensonal real representation, then 
\begin{equation}
w_2(R\otimes L)=w_2(R).
\end{equation}

In Section \ref{2d}, we propose that given a gapped 2d band Hamiltonian, the invariant $\beta\in H^2(G,\ZZ_2)$ of 2d fermionic SRE phases with symmetry $G\times\ZZ_2^F$ is given either by $w_2(R)$ or $w_2(R)+w_1(R)^2$, where $R$ is a certain representation of $G$. The supercohomology equation implies that $\beta\cup\beta\in H^4(G,\ZZ_2)$ maps to a trivial class in $H^4(G,U(1))$. To show that this is automatically the case for our two candidates, we note that for finite groups $H^4(G,U(1))\simeq H^5(G,\ZZ)$. The class in $H^5(G,\ZZ)$ corresponding to $\beta\cup\beta$ can be obtained by applying the Bockstein homomorphism $H^4(G,\ZZ_2)\ra H^5(G,\ZZ)$. A mod-2 class is annihilated by the Bockstein homomorphism if and only if it is a mod-2 reduction of an integral class. Now recall the well-known relation between Stiefel-Whitney classes and Pontryagin classes \cite{MilnorStasheff}:
\begin{equation}\label{MSrelation}
w_2^2= p_1\ {\rm mod}\, 2. 
\end{equation}
Hence $w_2^2$ is indeed annihilated by the Bockstein homomorphism. The same is true if we replace $w_2$ with $w_2+w_1^2$. Indeed, since
\begin{equation}
(w_2+w_1^2)^2=w_2^2+w_1^4,
\end{equation}
it is sufficient to show that $w_1^4$ maps to a trivial class in $H^4(G,U(1))$. Now we recall that $w_1^2$ is cohomologous to $\delta \omega/2$, where $\omega$ is an integral lift of $w_1$. Therefore $w_1^2$ is cohomologous to $\frac12\delta\omega\cup\frac12\delta\omega$, which is a coboundary of $\frac{1}{4}\omega\cup\delta\omega$. 

In Section \ref{1d}, we show that for a band Hamiltonian, the invariant $\halpha\in H^2(\hG,U(1))$ of 1d fermionic SRE phases with symmetry $\hG$ is equal to the image of $w_2(R)$ under the map $\iota: H^2(\hG,\ZZ_2)\ra H^2(\hG,U(1))$, for a particular representation $R$. Obviously, any element in the image of $\iota$ has order $2$, so in general not every element in $H^2(\hG,U(1))$ can be realized by a band Hamiltonian. But we claimed that for some $\hG$, even certain elements of order $2$ in $H^2(\hG,U(1))$ cannot be realized by band Hamiltonians. This happens because  not every element in $H^2(\hG,\ZZ_2)$ arises as $w_2(R)$ for some representation $R$.  The reason is again the relation (\ref{MSrelation}). It implies that for any representation $R$ of $\hG$, the Bockstein homomorphism annihilates $w_2(R)^2$. On the other hand, a generic element of $H^2(\hG,\ZZ_2)$ need not have this property. An example of a finite group $\hG$ for which some elements of $H^2(\hG,\ZZ_2)$ do not arise as $w_2(R)$ for any $R$ is given in \cite{metacyclic}.

\section{Beta as a charge pumping invariant}\label{betapump}

As discussed in Section \ref{1d}, fermionic SRE phases in 1d with symmetry $\hG$ have an invariant $\beta\in H^1(G,\ZZ_2)$. More precisely, this invariant is defined if the invariant $\gamma$ (the number of boundary fermionic zero modes modulo $2$) vanishes. The definition of $\beta$ given in Ref. \cite{FidKit} relies on the properties of boundary zero modes. Namely, $\beta(g)=1$ (resp. $\beta(g)=0$) if $g\in G$ acts on the boundary Hilbert space by a fermionic (resp. bosonic) operator. Here we explain an alternative formulation of $\beta\in H^1(G, \ZZ_2)$ as a charge pumping invariant. Any symmetry $\hg\in\hG$ gives rise to a loop in the  space of 1d band Hamiltonians. The net fermion parity pumped through any point is a $\ZZ_2$-valued invariant of the loop. This is a special case of the Thouless pump \cite{TeoKane10,MooreBalents}.

Given $\hg\in\hG$ which is a symmetry of a band Hamiltonian $H(k)$, we can define a loop in the space of band Hamiltonians as follows. Since $SO(2N)$ is a connected group, we can choose a path $\eta: [0,1]\ra SO(2N)$ such that $\eta(0)=1$ and $\eta(1)=\hR(\hg)$. Next we define $H(k,t)=\eta(t) H(k) \eta(t)^{-1}$. Since $\hR(\hg)$ commutes with $H(k)$, $H(k,1)=H(k,0)$. Thus $H(k,t)$ is a loop in the space of 1d band Hamiltonians. A general argument \cite{TeoKane10,MooreBalents} shows that the net fermion parity $(-1)^{B(\hg)}$ pumped through one cycle of this loop does not depend on the choice of path $\eta$. This immediately implies that $B(\hg\hg')=B(\hg)+B(\hg')$. Thus $B(\hg)$ defines an element of $H^1(\hG,\ZZ_2)$. 

To evaluate $B(\hg)$, we apply the general formula from Ref. \cite{TeoKane10} for Hamiltonians in class D. One simplification is that locally in $k,t$ the Berry connection can be taken as $\eta^{-1} \partial_t\eta$, and thus its curvature vanishes. Then
\begin{equation}\label{B}
B(\hg)=\frac{1}{2\pi}\int\Tr\left[(P_+(0)-P_+(\pi))\eta(t)^{-1}\partial_t\eta(t)\right]\,dt
\end{equation}
where $P_+(k)$ is the projector to positive-energy states at momentum $k$.

Next we decompose $\hR$ into real irreducible representations $r_\alpha$. Obviously, each representation contributes independently to $B(\hg)$. Representations of $\CC$-type and $\HH$-type do not contribute at all, since the corresponding Hamiltonians can be deformed  to trivial ones. A Hamiltonian $A_{r,ij}$ corresponding to an $\RR$-type representation $r_\alpha$ is of class D and can be deformed either to a trivial one or to a trivial one stacked with a single Kitaev chain. In the former case, both the boundary invariant $(-1)^{\beta(\hg)}$ and the charge-pumping invariant $B(\hg)$ are trivial (equal to $1$). In the latter case, we get a single Majorana zero mode for each of the $d_r=\dimr$ basis vectors of $r$, so the  boundary invariant $(-1)^{\beta(\hg)}$ is equal to $\det r(g)$. We just need to verify that $B(\hg)$ is also equal to $\det r(g)$ for $d_r$ copies of the Kitaev chain. The on-site representation of $\hG$ is given by $\hR=r\oplus r$ in this case.

For $d_r$ copies of the Kitaev chain, the projector to positive-energy states is
\begin{equation}
	P_+(k)=\tfrac{1}{2}\left(\mathds{1}_2-\sigma_y \sin k+\sigma_z \cos k\right)\otimes\mathds{1}_{d_r},
\end{equation}
which commutes with $\hR(\hg)=\mathds{1}_2\otimes r(\hg)$ and satisfies $P_+(0)\hR(\hg)=r(\hg)\oplus 0$ and $P_+(\pi)\hR(\hg)=0\oplus r(\hg)$. Let $\eta(t)$ be a path in $SO(2d_r)$ from $1$ to $\hR(\hg)$. We may choose it to belong to the $U(d_r)$ subgroup of matrices that commute with $P_+(0)$ and $P_+(\pi)$. Then $\eta(t)=q(t)\oplus\bar q(t)$ for a path $q(t)$ through $U(d_r)$ from $\mathds{1}$ to $r(\hg)$.

Substituting all this into (\ref{B}), we get 
\begin{equation}
B(\hg)=\frac{1}{2\pi}\int\Tr((P_+(0)-P_+(\pi))\eta(t)^{-1}\partial_t\eta(t))\,dt=\frac{1}{2\pi}\int\Tr(q(t)^{-1}\partial_tq(t)-\bar q(t)^{-1}\partial_t\bar q(t))\,dt.
\end{equation}
Note that this vanishes whenever $q(t)=\bar q(t)$ at all $t$. We now show how to recover $(-1)^{B(\hg)}=\det r(\hg)$.

If $r(\hg)$ has determinant $+1$, it lives in $SO(d_r)$, which is path-connected. Hence the path $q(t)$ from $\mathds{1}$ to $r(\hg)$ may be taken to lie in $SO(d_r)\subset U(d_r)$. Therefore $q(t)=\bar q(t)$ is real, and so $B(\hg)=0$.

If $r(\hg)$ has determinant $-1$, we construct $q(t)$ as follows. First connect $\mathds{1}$ to $\text{diag}(-1,+1,+1,\ldots,+1)$ by $\text{diag}(\exp(it),+1,+1,\ldots,+1)$. Now that the determinant is $-1$, we may get to $r(\hg)$ through a real path in the identity-disconnected component of $O(d_r)$. This second segment of the path contributes nothing to $B(\hg)$. It remains to compute the contribution of the first segment, where $q(t)=\exp(it)\oplus\mathds{1}$:
\begin{equation}
B(\hg)=\frac{1}{2\pi}\int\left(e^{-it}\partial_te^{it}-e^{it}\partial_te^{-it}\right)\,dt=1.
\end{equation}
This completes the proof that $B(\hg)=\beta(\hg)$. In particular, $B(P)=0$, i.e. $B$ is really a homomorphism from $G=\hG/\ZZ_2^F$ to $\ZZ_2$.

The interpretation of $\beta(g)$ in terms of a fermion-parity pump has the following intuitive reason. Assume that one can make a ``Wick rotation'' of the pump. Then the twist by $\hg$ along the ``time'' direction gets reinterpreted as a twist along the spatial direction. The invariant $B(\hg)$ can be re-interpreted as the fermionic parity of the ground state of the system with an $\hg$-twist, or equivalently as the fermionic parity of the $\hg$ domain wall. On the other hand, it is known \cite{KTY2} that this is yet another interpretation of the invariant $\beta$. 

To conclude this section, we show how to compute $B(g)=\beta(g)$ from the holonomy of the Berry connection between $k=0$ and $k=\pi$. This makes the topological nature of $B(g)$ explicit. Recall first how the holonomy is defined. If there are $2N$ Majorana fermions per site, a free 1d Hamiltonian can be described by a non-degenerate $2N\times 2N$ matrix $X(k)$, where $k$ is the momentum \cite{Ryuetal}. At $k=0$ and $k=\pi$ this matrix is real and skew-symmetric. We can bring $X(0)$ to the standard form $X_0$ using an orthogonal transformation $O(0)\in O(2N)$. Similarly, we use  $X(\pi)$ to define $O(\pi)\in O(2N).$ The holonomy of the Berry connection is $O=O(\pi) O(0)^{-1}$. The invariant $(-1)^\gamma$ is equal to the sign of $\det O$ \cite{gammadef}. If $\gamma$ vanishes, then $\det O(0)$ and $\det O(\pi)$ have the same sign, and by a choice of basis we may assume that both $O(\pi)$ and $O(0)$ lie in $SO(2N)$. 

To define a topological invariant associated to an element $\hg\in \hG$, we choose a path $\eta(t):[0,1]\ra SO(2N)$ from the identity to $\hR(\hg)$. Consider now the following map from $[0,1]$ to $SO(2N)$:
\begin{equation}
\Pi(t)=\left\{ \begin{array}{ll} \eta(2t), &  0\leq t\leq 1/2,\\
O \eta(2-2t) O^{-1}, & 1/2\leq t\leq 1.\end{array}\right .
\end{equation}
Since $O \equiv O(\pi)O(0)^{-1}$ is the holonomy of the Berry connection from $0$ to $\pi$, it commutes with all symmetries of the Hamiltonian, including $\hR(\hg)$ for all $\hg\in\hG$. This implies that $\Pi(t)$ is a loop in $SO(2N)$.
We claim that $B(\hg)$ is the class of this loop in $\pi_1(SO(2N)) = \ZZ_2$.

This definition is independent of the path from $1$ to $\hR(\hg)$. Any two paths differ (in the sense of homotopy theory) by a loop in $SO(2N)$. Thus changing the path  will result in composing $\Pi(t)$ with a loop and its conjugation by $O$. Since these two loops are homotopic, the homotopy class $\left[\Pi\right]$ is unchanged. 

To prove that the homotopy class of the loop $\Pi$ coincides with $B(\hg)$, one can follow the same strategy as before: use homotopy-invariance to reduce to the case of a single Kitaev chain, and then compute the invariant by choosing a particularly convenient path.

\end{document}